# Local Density Approximation Description of Electronic Properties of Wurtzite Cadmium Sulfide (w-CdS)

E. C. Ekuma[1], L. Franklin, G. L. Zhao, J. T. Wang, and D. Bagayoko

Department of Physics

Southern University and A&M College

Baton Rouge, Louisiana 70813, USA

**Abstract**

We present calculated, electronic and related properties of wurtzite cadmium sulfide (w-CdS). Our ab-initio, non-relativistic calculations employed a local density functional approximation (LDA) potential and the linear combination of atomic orbitals (LCAO). Following the Bagayoko, Zhao, and Williams (BZW) method, we solved self-consistently both the Kohn-Sham equation and the equation giving the ground state density in terms of the wave functions of the occupied states. Our calculated, direct band gap of 2.47 eV, at the $\Gamma$ point, is in excellent agreement with experiment. So are the calculated density of states and the electron effective mass. In particular, our results reproduce the peaks in the conduction band density of states, within the experimental uncertainties.

[1] Current Address: Department of Physics and Astronomy, Louisiana State University, Baton Rouge, LA 70803, USA

## I. Introduction and Motivations

Wurtzite cadmium sulfide (w-CdS) has long been recognized as an important optoelectronic, piezo-electronic, and semiconducting material [1]. In particular, thin films of CdS are of great interest due to their efficient utilization in the fabrication of solar cells.[1] Photovoltaic effects have been demonstrated in the heterojunction Cu-CdS at photon energies above and below the band gap, [2] where electron emission by Cu explains the effect for photon energies below the band gap. Lami and Hirlimann [3] obtained stimulated emission of green light (2.45 eV) from CdS at room temperature.

Basic, electronic properties of CdS have been experimentally established using different measurement techniques [1,3-8]. Photoemission measurements of Kindig and Spicer [4] provided details on the density of states of w-CdS for both the valence and the conduction bands while those of Stoffel [6] addressed band widths and other properties. The reported valence band width of 4.5 $\pm 0.2$ eV [6] corroborated the finding of Stoffel and Margaritondo [7]. Kingston et al. [8] provided detailed edge emission bands for bulk w-CdS at very low temperatures. Shin et al. [9] reported a low temperature optical band gap of 2.55 eV for w-CdS. The transmission edge and photocurrent study of these authors employed high purity w-CdS single crystal platelets, grown by sublimation method. Magnusson et al. [10] quoted a band gap of 2.58 eV for lattice constants of 4.13 and 6.70 Å, for “a” and “c,” respectively. The angle resolved inverse photoelectron spectroscopy of these authors found conduction band critical points at 5.8 eV and 7.5 eV, at the M point. Reference manuals [11,12] provide additional experimental data on wurtzite CdS. Chiang and Himpsel reported an electron effective mass of 0.21, at the conduction band minimum, in the direction perpendicular to the c axis.

Absorption measurements by Oliva et al. [13] found room temperature optical band gaps of 2.42 eV for polycrystalline thin films of CdS grown by chemical bath deposition and close spaced sublimation. Using chemical bath deposition, Ortuno-Lopez et al. [14] grew films of w-CdS in such a way that they could tune their band gaps. The decrease of the strain between the film and the glass substrate was found to decrease the band gap. The range of the values of the measured, optical band gaps of the resulting films was from 2.26 to 2.5 eV.

In contrast to the experimental studies, the theoretical results appear to cover a wide range of values for the band gap and other quantities. The empirical pseudopotential calculation of Bergstresser and Cohen [15] was one of the first theoretical studies of the band structure of wurtzite CdS (w-CdS). One of the earlier local density functional calculations, by Chang et al. [16], led to a band gap of 1.77 eV for w-CdS. These authors also reported the upper p valence band and the total valence band widths of 4.1 and 11.69 eV, respectively. A decade later, Schröer et al. [17,18] and Xu and Ching [19] reported extensive, calculated results for CdS. At calculated lattice constants of 4.03 and 6.54 Å, the LDA pseudopotential calculation of Schröer et al. [17] found a band gap of 1.31 eV. These authors employed non-local, separable and norm-conserving pseudopotentials and explicitly treated the d electrons as valence electrons. In their linear combination of Gaussian orbitals (LCGO) calculations, Xu and Ching [19] augmented their local density potential with an additional exchange obtained with Wigner interpolation. At experimental lattice constants of 4.137 and 6.7144 Å, with a value of 0.375 for the parameter u, Xu and Ching reported a band gap of 2.02 eV. These authors obtained electron effective masses of 0.27, 0.21, and 0.25 in the $\Gamma$- K, $\Gamma$- A, and $\Gamma$- M directions, respectively. Xu and Ching [19] also found 4.43 and 12.25 eV for the upper p valence and the total valence band widths. They discussed other calculated properties, including the dielectric function.

From the mid 1990s to present, several other theoretical reports [20-24] followed the above ones. Using pseudopotentials, Vogel et al. [21] found w-CdS band gaps of 1.2, 2.4, and 2.5 eV with LDA, self-interaction corrected, and self-interaction corrected with relaxation calculations. The results of Sharma et al. [22] underestimated the band gap by 1.75 and 0.5 eV with LDA and exact exchange potentials, respectively. They overestimated it by 1.5 and 1 eV with full potential and exact exchange calculations without and with core-valence interactions, respectively. The generalized gradient approximation calculations of Huang et al. [23] obtained 1.69 eV for the band gap of bulk w-CdS and 2.25 eV for CdS nanowires. With a modified version of the Becke and Johnson exchange potential, Tran and Blaha [24] found a gap of 2.66 eV for w-CdS. Their LDA result was 0.86 eV.

The density functional theory calculated band gaps for w-CdS cover a wide range from 0.75 (i.e., 2.5-1.75) and 0.86 eV to 2.66 eV. The disagreement between these theoretical values and between them and experiment, particularly for the band gap, is a key motivation for this work. This motivation is reinforced by the fact that quasiparticle calculations have not totally resolved the inability of theory to obtain, unambiguously, measured electronic properties of w-CdS.

Indeed, while the LDA calculations of Zakharov et al. [25] led to a gap of 1.36 eV, their quasiparticle calculations reported 2.79 eV, higher than the experimental value. Shishkin and Kresse [26] found 2.06, 2.26, and 2.55 eV from their $G_0W_0$, $GW_0$, and GW calculations, respectively. These authors also reported a generalized gradient approximation (GGA) calculation result of 1.14 eV. Fuchs et al. [27] obtained 2.55, 2.65, and 2.80 eV from their $G_0W_0$, $GW_0$, and GW calculations, respectively. These results suffice to see that quasiparticle calculations have not resolved the under- or overestimation of the band gap by theory.

In light of the preceding overview, the aim of this work is to attempt to obtain the measured value of the band gap and of other electronic properties of w-CdS with ab-initio, self-consistent LDA calculations. The confirmation [28-29] of our predictions of the band gaps and other properties for c-$Si_3N_4$ [30] and c-InN [31-32] is a basis for the above presumption. Further, the mathematical rigor of the method [32-34] suffices to expect it to lead to much better results.

## II. Method

Our calculations employed the Ceperley and Alder [35] local density functional potential as parametrized by Vosko, Wilk, and Nusair [36] and the linear combination of atomic orbitals. The radial parts of these orbitals are Gaussian functions. We utilized a program package developed and refined over decades [35,38]. Our calculations are non-relativistic and are performed at zero temperature. The distinctive feature of our approach resides in our implementation of the Bagayoko,

Zhao, and Williams (BZW) method consisting of concomitantly solving self-consistently two coupled equations. One of these equations is the Schrödinger type equation of Kohn and Sham [39], referred to as the Kohn-Sham (KS) equation. The second equation, which can be thought of as a constraint on the KS equation, is the one giving the ground state charge density in terms of the wave functions of the occupied states.

The essentials of the method follow. Beginning with a relatively small basis set capable of accounting for all the electrons in the system under study, one performs ab-initio, self-consistent calculations. Subsequently, one performs a second calculation with a larger basis set that includes that of Calculation I. The occupied energies levels of the two calculations are compared numerically and graphically. These occupied energies from Calculations I and II are generally different. A third calculation is carried out with a larger basis set that includes that of Calculation II. Again, the occupied energies from Calculation II and III are compared. This process continues until the occupied energies from a calculation, i.e., N, are found to be identical to those from Calculation (N+1), within our computational uncertainties of 5 meV or less. At that point, the calculations are completed and the results from Calculation N represent the physical description of the system under study. We have explained elsewhere [33,40] how to increase methodically the basis set from one calculation to the following. Calculation (N+1) and others with larger basis sets, provided linear dependency is avoided, reproduce the occupied energies from Calculation N. However, by virtue of the Rayleigh theorem, these calculations produce some unoccupied eigenvalues that are lower than those from Calculation N, due to a non-trivial basis set and variational effect [33,40]. It should be noted that the iterations for the KS equation are embedded in those of the charge density equation. Also, the convergence of the iterations for the charge density equation consists of reaching the minimum values for the occupied energies in going from a calculation to the next one. When this occurs, as shown in the comparison of occupied energies from Calculations N and (N+1), the basis set of Calculation N is referred to as *the optimal basis* set which leads to the optimal density for the actual ground state of the system.

The details needed to replicate our calculations follow. Wurtzite CdS possesses a hexagonal lattice in the space group $C^4_{6v}$. There are four atoms per unit cell, in the positions indicated between parentheses: Cd: (0, 0, 0), (1/3, 2/3, 1/2); S: (0, 0, u), (1/3, 2/3, 1/2 + u). Our self-consistent computations were performed at the experimental lattice constants of 4.136 Å and 6.713 Å, for “a” and “c,” respectively, with a “u” parameter of 0.37715. As in previous works, [40] we employed a mesh of 24 k-points, with proper weights, in the irreducible Brillouin zone. Our criterion for self-consistency of the iterative solutions of the KS equation consisted of the convergence of the potential to a difference around $10^{-5}$ between two consecutive iterations. Approximately 60 iterations were needed to reach self-consistency. The computational error for the valence charge was about 0.00841 for 88 electrons, a little less than $10^{-5}$ per electron.

Our computations started with those for the neutral Cd and S atoms. The wave functions from these calculations, for the occupied states, were utilized to perform a self-consistent calculation for w-CdS. Using the outputs of this trial calculation we estimated a charge transfer of approximately 2 electrons from Cd to S. We then performed ab-initio, self-consistent calculations for $Cd^{+2}$ and $S^{-2}$. The wave functions resulting from these calculations were utilized to construct the trial basis sets for the solid state calculations which started with a small basis set, one that is just large enough to account for all the electrons in the system. In this first calculation, the description of the valence states required (3d4s4p4d5s) orbitals on $Cd^{+2}$ and (2s2p3s3p) orbitals on $S^{-2}$. Calculation II employed the above basis set plus 5p orbitals on cadmium. A comparison of the occupied energies from Calculations I and II showed that they were clearly different. So, as per the BZW method, we performed other calculations. Calculation III added the 6s orbital on cadmium while Calculation IV further augmented the basis set on cadmium with the 5d orbitals. Calculation V increased the basis set on sulfur with 4p orbitals. Calculation VI added the 4s orbital to those on sulfur. The comparison of the occupied energies from Calculation V and VI showed that they were equal, within the computational uncertainties, signifying that the minimum occupied energies were reached by Calculation V. As noted above, the attainment of their minima by the occupied energies is the criterion for the convergence of the size of the basis set for the description of the *ground state*. Otherwise stated, this attainment marks the self-consistence of the “solution” of the charge density equation: out of the practically infinite number of possible trial basis sets, this attainment unambiguously determines *the optimal basis* set that describes the ground state charge density, following the embedded self-consistency of the solution of the KS equation. The results discussed below for w-CdS are those obtained from Calculation V, with the optimal basis set.

## III. Results

Our calculated, ab-initio, self-consistent bands for w-CdS are in Figure (1). They resulted from Calculation V, as explained above. As per the following comparison with experiments, these bands reproduce most experimental results not only for the valence, but also the conduction bands. The valence band maximum at the gamma point is a doublet that is 0.012 eV above the singlet. The crystal field splitting between the two is reported by Chiang and Himpsel[12] to be 0.015 eV, in basic agreement with our result. The calculated width of the group of upper valence p bands is 4.22 eV, much closer to the experimental value of 4.5 $\pm$ 0.2 eV [6] than most of the previous LDA and some other theoretical results. The reported, experimental width [6] of the low laying Cadmium 4d valence bands of 1.2 $\pm$ 0.2 eV is basically the same as the calculated one of 1 eV, within the experimental uncertainties. While experiment places these 4d bands at 9.4 $\pm$ 0.5 eV [4], our calculated minimum of these bands, at Gamma, is 8.33 eV. Stoffel[6] found the lowest valence band energy of -12.5 $\pm$ 0.2 eV. Our calculated minimum for these bands is -12.11 eV, at the $\Gamma$ point.

Table I provides a detailed comparison between measured valence band eigenvalues [6] and results from our Calculation V. Out of 16 experimental values in the table, our results agree with 10, within the experimental uncertainties. For the remaining 6 for which the calculated values deviate from measured ones, they do so by at most 5%. We predict two eigenvalues for which Stoffel [6] did not report a value. Magnusson et al. [10] identified critical energies in the conduction bands, at the M point, of 5.8 eV and 7.5 eV. In Figure (1), we have respectively 5.81 eV and 7.18 eV at the M point.

Figures (2) and (3) show the calculated, total (DOS) and partial (pDOS) density of states. Kindig et al. [4] reported a peak at -1.2 $\pm$ 0.3 eV in the valence bands DOS. The corresponding, calculated peak is at -1.3 eV, in excellent agreement with experiment. We predict a second, narrow, peak at -3.75 eV for which we could not find an experimental value. The experimentally measured peaks in the conduction band DOS are at 4.4 $\pm$ 0.5 eV, 6.7 $\pm$ 0.3 eV, and 8.2 $\pm$0.3 eV [4]. The corresponding calculated peaks are at 5 eV, 6.3-6.7 eV, and 8 eV, respectively, in excellent agreement with experiment within the experimental uncertainties. A broad structure in the calculated conduction band DOS, between 6.25 eV and 7 eV has a narrow peak around 6.3 eV and a broader one centered at 6.75 eV.

The calculated electron effective masses in the Γ-A, Γ–K, and Γ-M directions are 0.21, 0.29, and 0.29 $m_0$, respectively. The apparent anisotropy in the effective mass is expected in a wurtzite structure. The effective mass is a measure of the curvature of the calculated bands. The agreement between calculated and measured effective masses indicates an accurate determination of the shape of the bands. A reference manual [12] cited an experimental value of 0.21 for the electron effective mass in w-CdS, along the c axis (i.e., Γ-A direction). We could not find other values of the electron effective mass in w-CdS.

## IV. Discussions

The above agreements between our LDA-BZW results and experimental ones indicate the accuracy of density functional theory description of w-CdS, provided that one looks for and obtains an optimal basis set that is verifiably complete for the description of the *ground state*, on the one hand, and that is not unduly large or overcomplete, on the other hand. Hence, these discussions focus on an attempt to explain the large discrepancies between many non-BZW DFT calculations and their disagreement with experiment.

This explanation mainly rests on direct implications of the Rayleigh theorem whose statement follows. Let an eigenvalue equation be solved with a basis set of N orbitals and (N+1) orbitals, where the N orbitals of the first calculation are augmented with an additional one for the second calculation. Let the eigenvalues of the two calculations be ordered from the lowest to the highest. Then, the Rayleigh theorem states that the eigenvalues from the two calculations satisfy the following inequality: $E_i^{(N+1)} \leq E_i^N$ for all $i \leq N$. Hence, upon an increase of the basis set, a given eigenvalue decreases or remains the same, the latter case corresponding to the situation where it has reached its minimum value, within computational uncertainties. This theorem, coupled with the fact that density functional theory utilizes only the wave functions of the occupied states in the search for a self-consistent solution of the Kohn-Sham equation, leads to the situation where, upon the convergence of the size of the basis set vis à vis the description of the ground state, some unoccupied energies continue to be lowered with increasing sizes of the basis set. As Zhao et al. [33] have shown, this lowering is not due to a physical interaction – given that the charge density, the potential, and the Hamiltonian do not change once the occupied energies reach their minimum values. It is this “extra” lowering of unoccupied energies that has been identified as a basis set and variational effect [33].

Table II shows the orbitals utilized in ten different calculations we carried out. As we have done for wurtzite ZnO [40], several of these calculations are partly intended to illustrate the fact that a single trial basis set calculation does not generally lead to a correct DFT description of non-metallic materials. Calculations VII and IX did not lower the occupied energies as compared to Calculation V. Specifically, the occupied bands from Calculation VII are slightly higher by about 30 meV than those of Calculations V and XI that are identical, as shown in Figure (4). Calculations VII and IX did not drastically change the lowest laying, unoccupied energies as compared to the Calculation V, In contrast, Calculations VIII and X did. Hence, for most practical purposes, Calculations VII and IX provide acceptable DFT descriptions of w-CdS. Indeed, the above difference of 30 meV is negligible in comparison to the above discrepancies between the band gaps from several non-BZW calculations and experimental findings. The band gaps resulting from VII and IX are respectively 2.52 eV and 2.48 eV. The closeness of the results from Calculations V, VII and IX indicates a robustness of the BZW method leading to providing a DFT description of w-CdS.

The main difference between Calculations VII and IX, on the one hand, and Calculations VIII and X, on the other, seems not to be the numeric sizes of the basis sets, but rather the inclusion of the sulfur 4s orbital in the basis sets for VIII and X. For most two or more atom systems, this behavior could understandably be ascribed to drastic deviations of the electronic cloud from spherical symmetry. These deviations are expected to be pronounced in solids. As stated elsewhere [33,40], the absence of a gap leads to the concomitantly convergence of the occupied and of the low laying unoccupied energies, explaining the early success of DFT and earlier $X_\alpha$ potentials in describing metals – provided one utilizes a large enough basis set to ensure its completeness for the description of the ground state.

## V. Conclusion

Our ab-initio, self-consistent LDA-BZW calculations led to electronic and related properties that mostly agree with experiment. Specifically, the calculated band gap of 2.47 eV is in accord with experiment. Our calculations reproduced measured peaks in the conduction band density of state. These agreements point to the accuracy of the density functional description of w-CdS, provided one utilizes a basis set that is complete for the description of the ground state and that is not overcomplete. The need for the BZW method in self-consistent calculations of electronic properties follows from the Rayleigh theorem, particularly for materials where an energy or band gap exists between occupied and empty states. With the method, the prospects seem great for DFT to inform and to guide the design and fabrication of semiconductor based devices. Further, theory could aid in the search for novel materials with desired properties, including binary to quaternary systems.

**Acknowledgments**: This work was funded in part by the Louisiana Optical Network Initiative (LONI, Award No. 2-10915), the Department of the Navy, Office of Naval Research (ONR, Award Nos. N00014-98-1-0748 and N00014-04-1-0587), the National Science Foundation (Award No. 0754821), and Ebonyi State, Federal Republic of Nigeria (Award No: EBSG/SSB/FSA/040/VOL. VIII/039).

**Table I.** Selected, calculated valence band eigenvalues [E(k) in eV] of w-CdS as compared to the experimental data of Stoffel.[6] Ten (10) calculated values agree with experiment, within the experimental uncertainties ($\pm 0.2$ eV). The six (6) values that fall outside the range of the experimental uncertainties do so at the percentage levels indicated between parentheses: $\Gamma_3$ (1.7%), the lowest $L_{1,3}$ (2.3%), the lowest $M_3$ (5%) and $M_1$ (3.3%), $A_{1,3}$ (1.9%), and the two $H_3$ (2.3%), and $H_3$ (4%).

| Symmetry Label | Measured E(k) | Calculated E(k) | Symmetry Label | Measured E(k) | Calculated E(k) | Symmetry Points | Measured E(k) | Calculated E(k) |
|---|---|---|---|---|---|---|---|---|
| $\Gamma_6$ | 0.0 | 0.0 | $M_4$ | -0.7 | -0.73 | $A_{5,6}$ | -0.5 | -0.38 |
| $\Gamma_1$ | ? | -0.012 | $M_3$ | -1.1 | -1.27 | $A_{1,3}$ | -2.6 | -2.35 |
| $\Gamma_5$ | -0.8 | -0.74 | $M_2$ | -1.7 | -1.85 | $H_3$ | -1.3 | -1.07 |
| $\Gamma_3$ | -4.5 | -4.22 | $M_1$ | ? | -2.59 | $H_{1,2}$ | -2.5 | -2.43 |
| $L_{2,4}$ | -1.4 | -1.27 | $M_3$ | -3.0 | -3.35 | $H_3$ | -4.2 | -3.83 |
| $L_{1,3}$ | -1.5 | -1.33 | $M_1$ | -4.3 | -3.96 | | | |
| $L_{1,3}$ | -4.3 | -4.0 | | | | | | |

**Table II**. Successively larger, trial basis sets, as per the BZW method, for the description of the valence states of wurtzite cadmium sulfide (w-CdS). The optimal basis set is that from Calculation V.

| **Calculation Number** | **Atomic Functions on Cd** | **Atomic Functions on S** | **Total Number of orbitals** | **Band Gap (eV)** |
|---|---|---|---|---|
| I | 3d4s4p4d5s | 2s2p3s3p | 46 | 3.1478 |
| II | 3d4s4p4d5s**5p** | 2s2p3s3p | 52 | 3.3740 |
| III | 3d4s4p4d5s5p**6s** | 2s2p3s3p | 54 | 2.7062 |
| IV | 3d4s4p4d5s5p6s**5d** | 2s2p3s3p | 64 | 2.6080 |
| **V** | **3d4s4p4d5s5p6s5d** | **2s2p3s3p4p** | **70** | **2.4727** |
| VI | 3d4s4p4d5s5p6s5d | 2s2p3s3p4p**4s** | 72 | 1.8039 |
| | | | | |
| VII | 3d4s4p4d5s5p6s5d | 2s2p3s3p3d | 74 | 2.5195 |
| VIII | 3d4s4p4d5s5p6s5d | 2s2p3s3p3d**4s** | 76 | 1.8542 |
| IX | 3d4s4p4d5s5p6s5d | 2s2p3s3p3d**4p** | 80 | 2.4771 |
| X | 3d4s4p4d5s5p6s5d | 2s2p3s3p3d4p**4s** | 82 | 1.8104 |

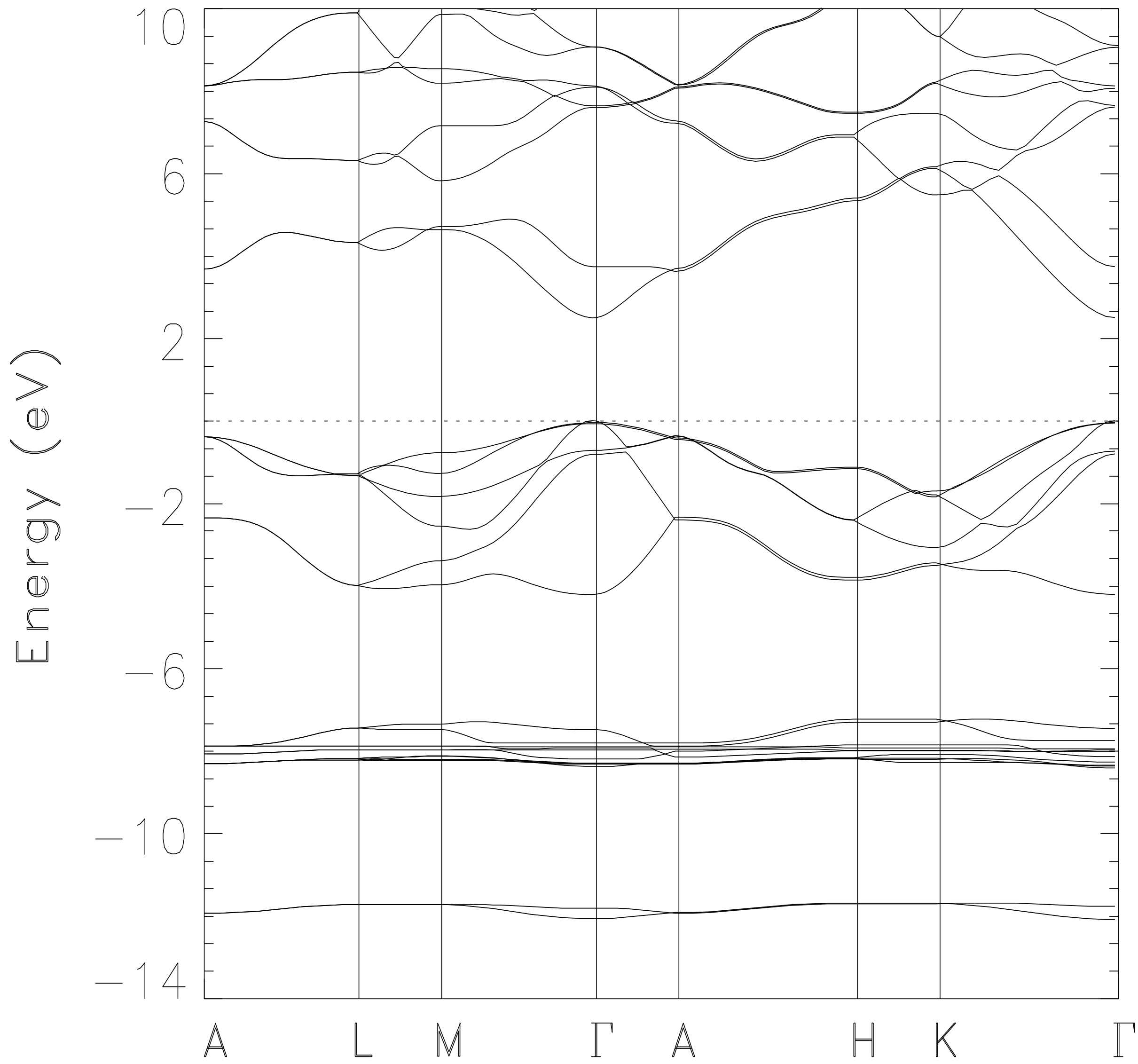

**Figure 1.** Electronic energy bands of w-CdS as obtained from Calculation V with the optimal basis set.

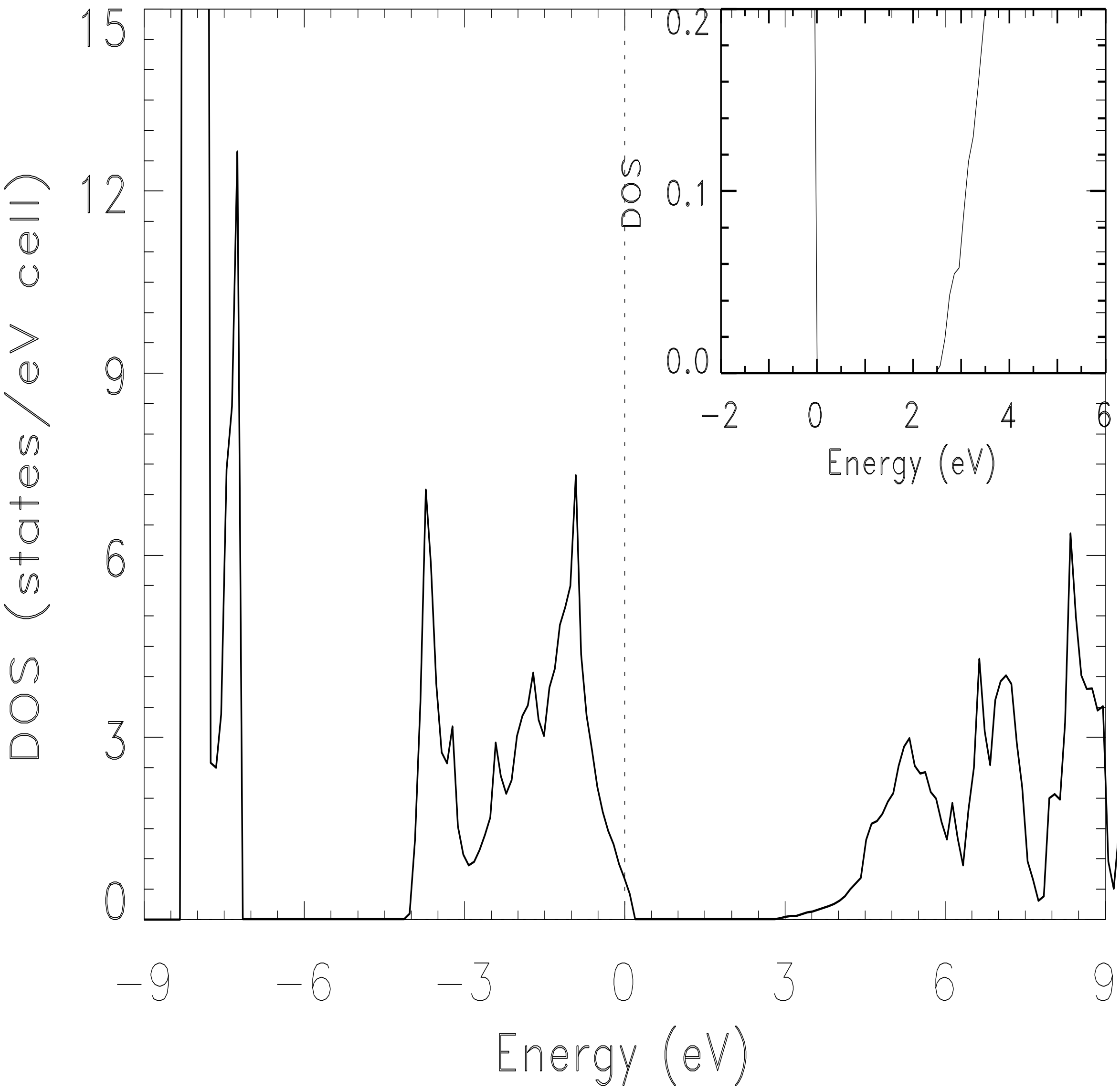

**Figure 2.** The total density of states (DOS) for w-CdS as obtained from the bands in Figure 1.

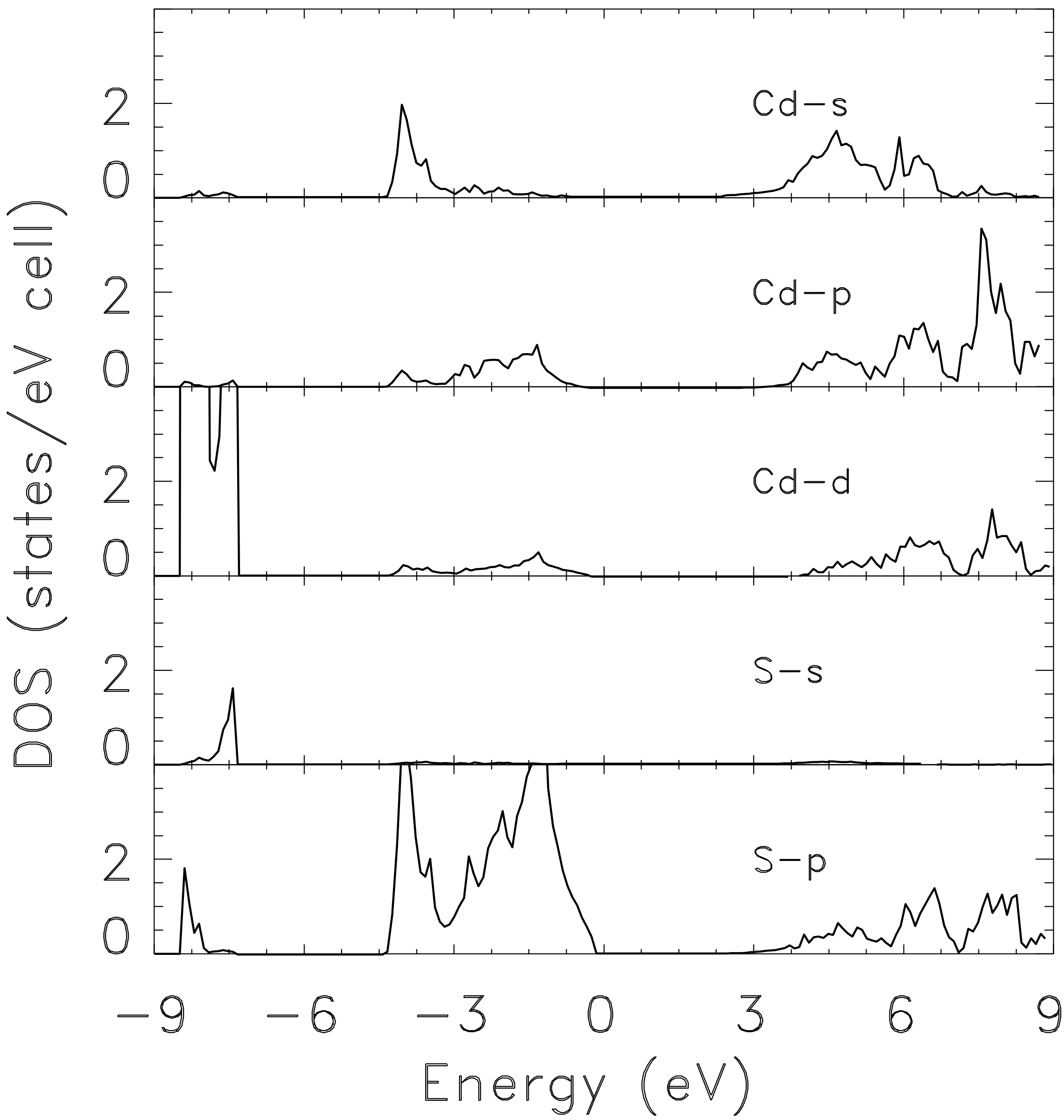

**Figure 3.** The partial densities of states of states (pDOS) for w-CdS as obtained from the bands in Figure 1.

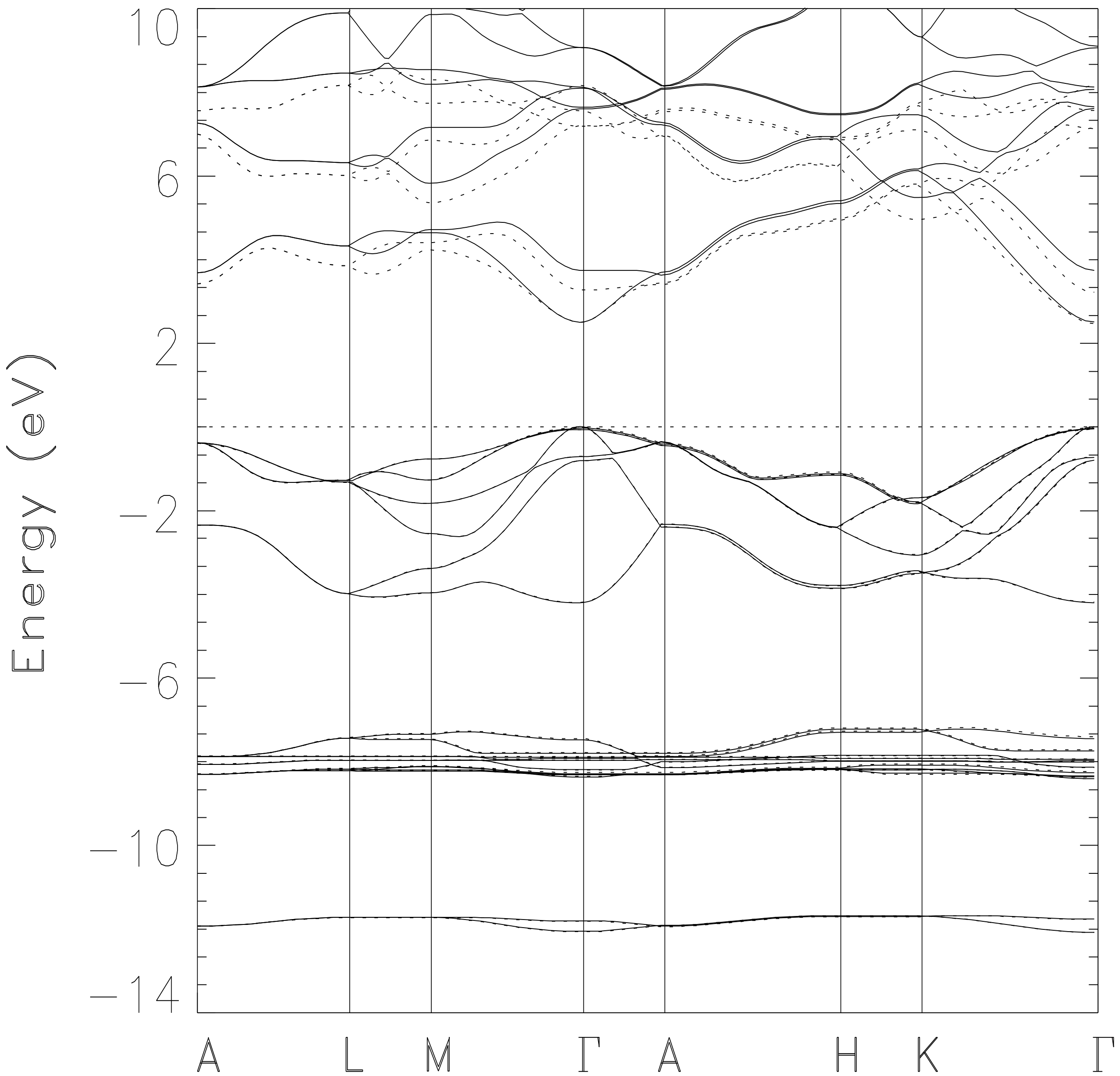

**Figure 4.** Electronic energy bands of w-CdS from Calculations V (full lines) and IX (dashed lines). The true DFT results, given the total convergence of the occupied states, are those of Calculation V.